\RequirePackage{fix-cm}
\documentclass[smallextended]{svjour3}       
\smartqed  
\usepackage{graphicx}
\newcommand{\cP}{{\cal P}}
\newcommand{\cT}{{\cal T}}

\newcommand{\cPT}{{\cal PT}}
\usepackage{epsfig}\usepackage{latexsym}\usepackage{amssymb}\usepackage{bm}
\usepackage{float}
\usepackage{subfigure}
\usepackage{amsmath}
\usepackage{amsfonts}
\usepackage[usenames]{color} 
\newcommand{\be}{\begin{equation}} \newcommand{\ed}{\end{displaymath}}
\newcommand{\bd}{\begin{displaymath}} \newcommand{\ee}{\end{equation}}
\newcommand{\bea}{\begin{eqnarray}} \newcommand{\eea}{\end{eqnarray}}
\newcommand{\ba}{\begin{array}} \newcommand{\ea}{\end{array}}

 \journalname{International Journal of Theoretical Physics}
\begin{document}

\title{Vector Models in $\cPT$ Quantum Mechanics}


\author{Katherine Jones-Smith      \and
        Rudolph Kalveks 
}


\institute{Katherine Jones-Smith \at
Physics Department\\
              Washington University in Saint Louis \\ 
             1 Brookings Drive\\
              Saint Louis, MO 63130 \\ 
              U.S.A.\\
              Tel.: +314-935-5162\\
              \email{kas59@physics.wustl.edu}           
           \and
           Rudolph Kalveks \at
           Theoretical Physics, Imperial College London\\
           London SW7 2AZ, U.K. 
}

\date{Received: 30 May 2012 / Accepted: 14 Jan 2013}


\maketitle

\begin{abstract}We present two examples of non-Hermitian Hamiltonians which consist of an unperturbed part plus a perturbation that behaves like a vector, in the framework of $\cPT$ quantum mechanics. The first example is a generalization of the recent work by Bender and Kalveks, wherein the E2 algebra was examined; here we consider the E3 algebra representing a particle on a sphere, and identify the critical value of coupling constant which marks the transition from real to imaginary eigenvalues. Next we analyze a model with SO(3) symmetry, and in the process extend the application of the Wigner-Eckart theorem to a  non-Hermitian setting.
\keywords{Non-Hermitian quantum mechanics \and PT quantum mechanics \and Wigner-Eckhart theorem}
\PACS{11.30.Er \and 03.65.-w \and 02.30.Fn}
\end{abstract}

\section{Introduction}
\label{intro}

There are many situations in quantum mechanics wherein the Hamiltonian under consideration can be written as
\be
 H=H_0+H_1
\label{genham}
\ee
where $H_0$ is the unperturbed part and commutes with the generators $T_i$ of symmetry group $G$:
\be
\ [H_0,T_i\ ]=0
\ee
and $H_1$ can be treated like a perturbation and behaves like a vector under $G$. 
We wish to generalize this situation in the context of $\cPT$ quantum mechanics  \cite{bender:1998ke,bender:1998gh}, where the assumption that operators such as the Hamiltonian are Hermitian is relaxed, and replaced by 
other requirements, notably that the Hamiltonian commutes with the parity ($\cP$) and time-reversal ($\cT$)
operators.

Interest in non-Hermitian quantum mechanics continues to grow \cite{bender:2007nj}, and recently a number of experiments have observed the so-called $\cPT$ phase transition, where the eigenvalues of a $\cPT$ Hamiltonian make a transition from being complex to real once a critical value of a coupling constant is reached \cite{guo:2009aa}, \cite{ruter:2010ce}, \cite{zhao:2010kf}. Thus it is relevant to seek new
$\cPT$-counterparts to conventional Hamiltonians. 

In this work we present two simple cases that can be described as non-Hermitian vector perturbation models where the Hamiltonian can be written as in eq (\ref{genham}); first we consider a particle confined to the surface of a sphere, where the Hamiltonian acts within an infinite dimensional Hilbert space, and next we consider a generic vector perturbation within a finite dimensional Hilbert space and determine the spectrum of eigenvalues using the Wigner-Eckart theorem. We find that for a range of parameters each of these models has a 
pure real spectrum. At critical values of the coupling the model undergoes $\cPT$ transitions wherein the eigenvalues become complex.

\section{E3 Algebra: particle on a sphere}
\label{sec:e3}
We begin by generalizing the analysis presented in \cite{benderkalveks}. They considered the E2 algebra which consists of elements $J,u,v$ such that 
\be
[J,u]=-iv \;\;\;\; [J,v] = iu \;\;\;\; [u,v]=0.
\ee
The Hamiltonian 
\be
h=J^2+igu
\label{e2ham}
\ee
where $J=-i\partial/\partial \theta$, $u=\sin\theta$, $v=\cos\theta$ and $g$ is a constant, represents a  2-dimensional quantum particle restricted to radius $r=1$. 

A generalization of this is the E3 algebra and restricting the particle to the surface of a sphere ($r=1$). This is described by the Hamiltonian 
\be
h=L^2 + igu_z
\label{e3ham}
\ee where $L$ obeys
\be
 [ L_i,L_j]=i
 \epsilon_{ijk}L_k
 \ee
 $u$ is a vector operator whose components are given by
\begin{align}
u_x&= \sin\theta\cos\phi \\
 u_y&=\sin\theta\sin\phi\\
 u_z&=\cos\theta
\end{align}
and $g$ is a constant. The remaining commutators are straightforward to calculate; 
\be
[L_i,u_j]=i\epsilon_{ijk}u_k \;\;\;\; [u_i,u_j]=0. 
\ee

Following
Bender and Kalveks we consider the case of even time reversal: for a wave function $\psi\left(\theta,\phi\right)$ the time reversal operator $\cT$ is manifested as complex conjugation:
\be
T\psi\left(\theta,\phi\right)=\psi^*\left(\theta,\phi\right)
\ee
hence $\cT^2=1$.
It is easy to verify the action of $\cT$ on the elements of the algebra: $\cT L_i \cT = -L_i$ and $\cT u_i \cT=u_i$.
The parity operator $\cP$ takes $\psi$ into the antipodal point:
\be
P\psi\left(\theta,\phi\right) = \psi\left(\pi-\theta,\phi+\pi\right)
\ee
so $\cP^2=1$; 
elements transform under parity as $\cP L_i \cP = L_i$ and 
$\cP u_i \cP=-u_i$. 
Note that
the Hamiltonian $h$ in eq (\ref{e3ham}) commutes with the combined operation $\cPT$ but not with $\cP$ or $\cT$ individually. 
Now let us determine the eigenvalue spectrum of this Hamiltonian. 
 We wish to solve
\be
h\psi\left(\theta,\phi\right)=\lambda\psi\left(\theta,\phi\right)
\ee
and we try the general solution:
\be
\psi\left(\theta,\phi\right)=f\left(\theta\right)e^{im\phi}.
\ee
For convenience we define $\eta=\textup{cos}\theta$; this simplifies the eigenvalue equation for $f$ : 
\be
-(1-\eta^2)\frac{\partial^2 f}{\partial \eta^2} +2\eta \frac{\partial f}{\partial \eta} + \frac{m^2}{1-\eta^2}f + ig\eta f = \lambda f
\ee
where $m$ is a fixed integer. 
If we let 
\be
h_0 = -(1-\eta^2)\frac{\partial^2 f}{\partial \eta^2} +2\eta \frac{\partial f}{\partial \eta} + \frac{m^2}{1-\eta^2}f
\ee
then the Hamiltonian we wish to solve is
\be
h_0f+ig\eta f = \lambda f.
\label{eq:newham}
\ee
We impose the boundary
condition that the solution must be regular at $\eta =  \pm 1$.

\begin{figure} 
 \includegraphics[width=6cm]{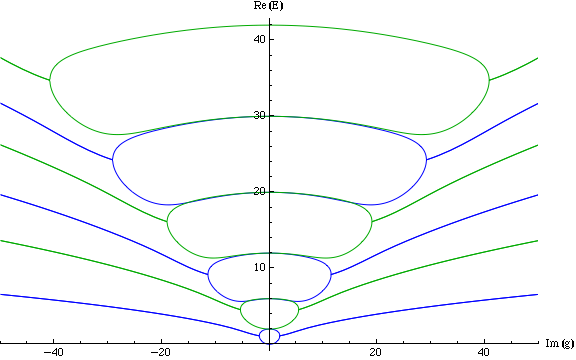}\;\;\;
 \includegraphics[width=6cm]{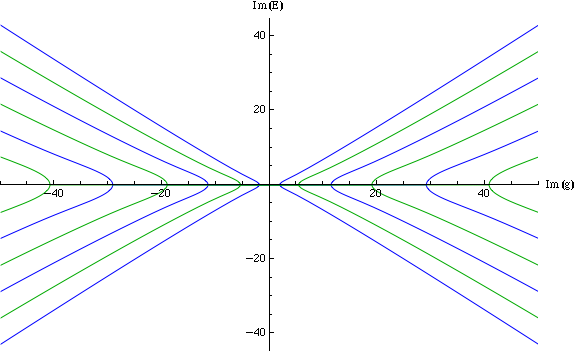}
\caption{Real and imaginary components of eigenvalues $E$ for the Hamiltonian given by Eq.\ref{eq:newham}. 
The first six eigenvalues for $m=0$ (blue) and  $m=1$ (green) are shown. Intercepts on the $E$ axis are given by $\ell(\ell+1)$ for $\ell=0 \;{\rm to}\; 6$.  For the case of $m=0$, we find that 
the spectrum is entirely real for $0 \leq g < 1.899$ at which point there is a transition to one
pair of complex conjugate eigenvalues in the spectrum. At $g = 11.45$ there is a second transition, to two pairs of 
complex conjugate eigenvalues. Similarly for the case
of $m=1$, we find one complex conjugate eigenvalue pair at 
$g=5.41$, and two pairs at $g=19.04$. In these 
computations
the Hamiltonian is truncated to a 100$\times$100 matrix;
we have verified that the relevant part of the spectrum is insensitive to the truncation.
}
 \label{figure}
\end{figure}

Let us choose basis elements 
\be
| l \rangle \rightarrow N_l P_{l,|m|}(\eta)
\ee
where $l=|m|, |m|+1,\ldots$, $P_{l,|m|}$ are the associated Legendre polynomials, with conventional normalization factor
\be
N_l = \sqrt{\frac{(2l+1)}{2}}\sqrt{\frac{(l-|m|)!}{(l+|m|)!}}.
\ee
The $P_{l,m}$'s satisfy
\be
h_0P_{l,|m|}(\eta)=l(l+1)P_{l,|m|}(\eta)
\ee
so the matrix of $h_0$ in this basis is diagonal. The matrix elements of the potential term, $i g \eta$, can easily
be determined from the normalization and recursion relations of the $P_{l,m}$'s. By diagonalizing the
truncated Hamiltonian matrix we can numerically obtain the eigenvalues of eq (\ref{eq:newham}); see
Fig.~(\ref{figure}).


\section{$\cPT$ Vector Model in Finite-Dimensional Hilbert Space}

$E3$ may also be regarded as a realization of the $\cPT$ vector model with
symmetry group $SO(3)$ and for which the Hilbert space is infinite dimensional. 
Now we wish to turn out attention to realizations of the $\cPT$ vector model
with finite dimensional Hilbert spaces. 
Let us write a simple, generic Hamiltonian $H=H_0 +H_I$ where 
\be
H_0=L_x^2+L_y^2+L_z^2
\ee
 \be
 H_I=V_z
 \ee
 and $V_z$ is the $z$ component of a vector operator. 
 
Our task is to obtain a matrix representation of the total Hamiltonian, solve for its eigenvalues and determine what value of the non-Hermitian perturbation cause the eigenvalues to become complex. 

Naturally we choose to work with the angular momentum eigenstates  $|\ell,m\rangle$; the action of $H_0$ on these states is well known, and we can utilize the Wigner-Eckart theorem to determine the action of $H_I=V_z$. 
  
Note that the dimensionality of the relevant vector space depends on the angular momenta of the multiplets but clearly it is finite. Suppose we consider the two multiplets
$|\ell,m\rangle$ and $|\ell+1,m\rangle$; 
$m$ takes on values from $-\ell$ to $+\ell$ in the first multiplet and from
$-\ell -1$ to $\ell + 1$ in the second multiplet,
so there are $(2\ell +1)+(2\ell+3)=4\ell + 4$ of these states. 

The action of $H_0$ on these states is simply 
\be
L^2|\ell,m\rangle=\ell(\ell+1)|\ell,m\rangle
\ee
\be
L^2|\ell+1,m\rangle=(\ell+1)(\ell+2)|\ell+1,m\rangle.
\ee
So all that remains is to determine how $V_z$ acts on these states; here we employ the Wigner-Eckart theorem, which we have extended to the non-Hermitian case as detailed in Appendix A. We find $\langle \ell^\prime,m^\prime|V_z|\ell,m\rangle=0$ unless $m=m^\prime$. Thus we need only to determine 
\begin{align*}
\langle \ell,m|V_z|\ell,m\rangle,\\
\langle \ell+1,m|V_z|\ell+1,m\rangle,\\
\langle \ell,m|V_z|\ell+1,m\rangle,\;{\rm and}\\
\langle \ell+1,m|V_z|\ell,m\rangle\\
\end{align*}
in order to completely specify $V_z$ in this space. The first two in this list can be expressed in terms of the reduced matrix element $\alpha$ defined in Appendix A; in general we find  
\begin{align*}
\langle \ell,m|V_z|\ell,m\rangle&=m \alpha_1\\
\langle \ell+1,m|V_z|\ell+1,m\rangle&=m\alpha_2;\\
\end{align*}
however we also wish to enforce $\cP V_z \cP=-V_z$ and $\cT V_z \cT=-V_z$, which restricts
$\alpha_1=\alpha_2=0$. (Determination of $\cP$ and $\cT$ within this space follows straightforwardly from their action on the spherical harmonics $PY_{\ell m}(\theta,\phi)=(-1)^\ell Y_{\ell m}(\theta,\phi)$ and 
$TY_{\ell m}(\theta,\phi)= Y^\ast_{\ell m}(\theta,\phi)=(-1)^m Y_{\ell, - m}(\theta,\phi)$.)

For the other two types of matrix elements, $\langle \ell,m|V_z|\ell+1,m\rangle$ and$\langle \ell+1,m|V_z|\ell,m\rangle$, we find these are proportional to other reduced matrix elements $\beta$ and $\gamma$;
\begin{align*}
\langle \ell+1,m|V_z|\ell,m\rangle&=f_{\ell m}\beta,\\
\langle \ell,m|V_z|\ell+1,m\rangle&=f_{\ell m}\gamma\\
\end{align*}
where 
\be
f_{lm}=\left[ \frac{(\ell+1)^2-m^2}{(2\ell+1)(2\ell+2)} \right]^{1/2}.
\ee 
Note that $f_{\ell m}$ is even in $m$.
 When we enforce $PV_zP=-V_z$ and $TV_zT=-V_z$, we find this requires $\beta$ and $\gamma$ to be pure imaginary, so we define $\beta=ib$, $\gamma=ic$ for some real numbers $b,c$. Note that in determining these matrix elements we do not assume $V$ is Hermitian; we rely only on the commutators of $V$ with the angular momentum operators. (See Appendix A for details.)

Now let us write down the matrix corresponding to the Hamiltonian $H = H_0 + H_I$. Consider the two-dimensional
subspace spanned by the states $| \ell, m \rangle$ and $| \ell + 1, m \rangle$ for a fixed value of $m$ that lies in the range $- \ell, \ldots, \ell$. Within this subspace
\be
H_0=\left(
\ba {cc}
\ell(\ell+1)&0\\
0&(\ell+1)(\ell+2)\\
\ea \right)
\ee
and
\be
V_z=\left(
\ba {cc}
0&-icf_{\ell m}\\
-ibf_{\ell m}&0\\
\ea \right).
\ee
In addition consider the two dimensional subspace spanned by the states $| \ell + 1, \ell + 1 \rangle$ and
$| \ell + 1, - \ell - 1 \rangle$. These states are not coupled by the perturbation $V_z$ to any other state and hence
$V_z = 0$ within this subspace. On the other hand the unperturbed Hamiltonian in this subspace is given by
\be
H_0=\left(
\ba {cc}
(\ell+1)(\ell+2)&0\\
0&(\ell+1)(\ell+2)\\
\ea \right).
\ee
It is convenient to define
\be
h_{\ell+1}=\left(
\ba {cc}
(\ell+1)(\ell+2)&0\\
0&(\ell+1)(\ell+2)
\ea \right) 
\ee
and
\be
h_{m}=\left(
\ba {cc}
\ell(\ell+1)&-icf_{lm}\\
-ibf_{lm}&(\ell+1)(\ell+2)
\ea \right) 
\ee
where $m = - \ell, \ldots, \ell$. 
The Hamiltonian can now be written as a block-diagonal matrix
\be
 \left( \ba {ccccc}
h_{\ell+1} & & & & \\
 &h_\ell& & & \\
& &h_{\ell-1}& &  \\
& && \rotatebox{-35}{\huge{...}}& \\
& & & & h_{-\ell}
\ea
\right).
\ee

\begin{figure} 
 \includegraphics[width=6cm]{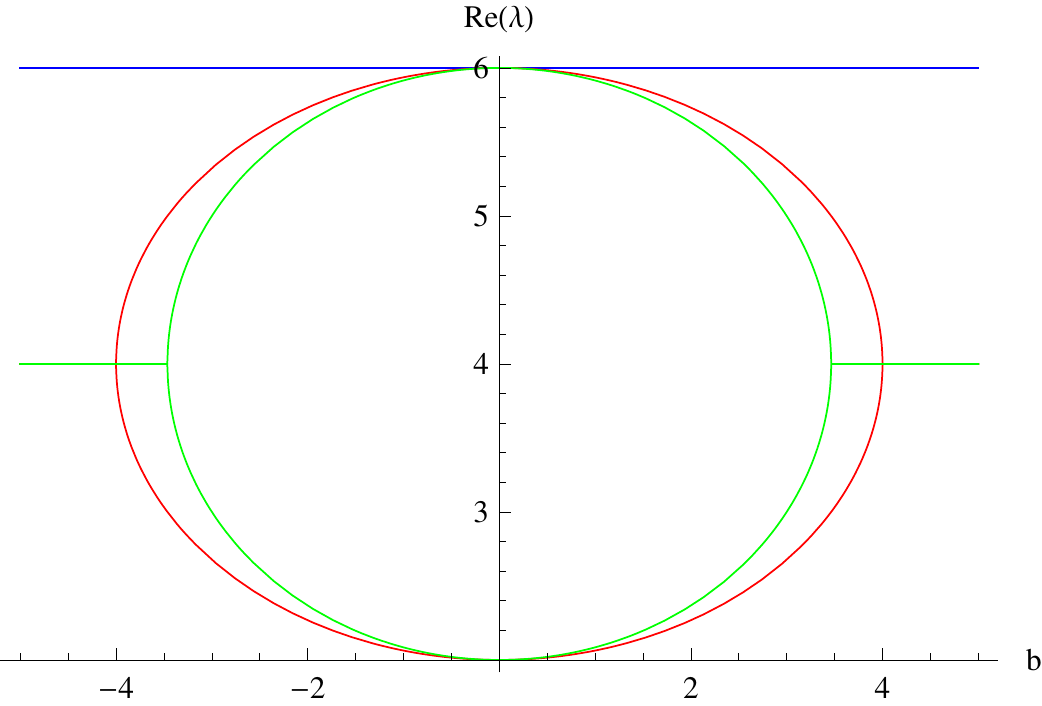}\;\;\;
 \includegraphics[width=6cm]{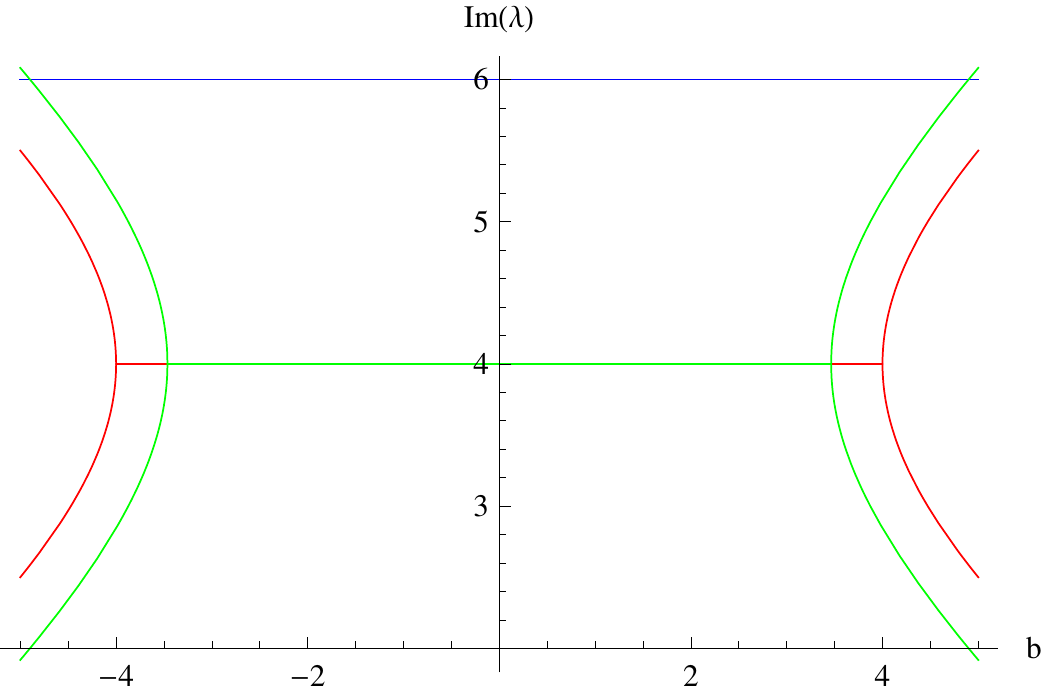}
 \caption{Real and imaginary parts of the eigenvalues $\lambda_{1,m}$ assuming $b=c$. The blue line at $\lambda=6$ corresponds to the $2\times2$ matrix denoted $h_{\ell+1}$ in the text, with eigenvalues $(\ell+1)(\ell+2)$. The other eigenvalues are $m$-dependent and correspond to the $2\times2$ matrices denoted $h_m$ in the text, with eigenvalues given by eq(\ref{lambdas}).  As noted $\lambda_{\ell,m}=\lambda_{\ell, -m}$, so there are only two distinct  $m$-dependent curves for $\ell=1$. In each figure, $|m|=1$ is plotted in red and $m=0$ is plotted in green.  Note that the transition to complex eigenvalues  occurs at $b=4$ for $\lambda^\pm_{1,1}$ and $b=\sqrt{12}\approx 3.47$ for $\lambda^\pm_{1,0}$.}
 \label{eigs}
\end{figure}

The individual $2\times2$ matrices that constitute the Hamiltonian are simple enough that we can obtain analytic expressions for the eigenvalues. The eigenvalues of $h_{\ell+1}$ are two-fold degenerate and are simply $(\ell+1)(\ell+2)$. The eigenvalues of $h_m$ are 
\be
\lambda_{\ell m}^{\pm}=(\ell+1)^2 \pm \sqrt{(\ell+1)^2-bcf_{\ell m}^2}
\label{lambdas}
\ee
Note that $\lambda_{\ell, m}=\lambda_{\ell, -m}$ so for all $m \neq 0$ the eigenvalues of $h_m$ are also two-fold degenerate. Clearly, the eigenvalues are real provided
\be
bc<\frac{(\ell+1)^2}{f_{lm}^2}.
\label{eq:ptcondition}
\ee
We can make the following observations about the behavior of the eigenvalues. Once $\cPT$ symmetry is broken, $\lambda^+$ and $\lambda^-$ form a complex conjugate pair.
Since $f_{\ell m}$ has its maximum value for $m=0$, $\lambda^{\pm}$ becomes complex for $m=0$ first. Similarly,  $f_{\ell m}$ is minimal for $|m|=\ell$, so $\lambda^{\pm}$ so these are the last eigenvalues to go complex.
For example we consider the case $\ell=1$ and choose $b=c$ for simplicity. We plot the eigenvalues in Figure \ref{eigs}.  

It is worth noting that in the Hermitian case $b = -c$. Hence the condition in 
eq (\ref{eq:ptcondition}) that ensures the eigenvalues are real is always met. 

\section{Conclusion}
We conclude by noting two natural generalizations of our results that deserve further
investigation. First the model of a particle on an ordinary 2-sphere considered in section II may be
generalized to a particle on a sphere in $n$ dimensions. The $\cPT$ transition for this model may be
amenable to analytic study in the large $n$ limit and may shed some light on $\cPT$ symmetric 
non-linear sigma models of which it would represent a $0+1$ dimensional case \cite{Coleman:1974jh}. Second the vector
model constructed in section III may be easily generalized from the symmetry group SO(3) to any Lie group
and therefore represents only one member of a large class of such models.

\newpage
\begin{center}
{\bf Appendix A: Wigner-Eckart Theorem}
\end{center}

Suppose we have an angular momentum operator ${\mathbf L}$ and a vector
operator ${\mathbf V}$ satisfying the commutation relations
\begin{equation}
[ L_i, V_j ] = i \epsilon_{ijk} V_k.
\label{eq:vectoralgebra}
\end{equation}
Let $| \ell, m \rangle$ denote an angular momentum multiplet of
total angular momentum $\ell$ and $z$-component $m$. Then
according to the Wigner-Eckart theorem the matrix
elements of $V_z$ and $V_\pm = V_x \pm i V_y$ between multiplet states
are determined by the commutation relations eq (\ref{eq:vectoralgebra}).
In the usual Wigner-Eckart theorem the Cartesian components of the operator
${\mathbf V}$ are assumed to be hermitian. Here we present a non-Hermitian
generalization of the theorem.

Following the usual arguments we find the selection rules
 \begin{align}
 \langle  \ell^\prime, m^\prime|V_z|\ell, m\rangle &=0 \;\;\;\;\;\; {\rm unless}\;m^\prime=m\\
  \langle  \ell^\prime, m^\prime|V_+|\ell, m\rangle &=0 \;\;\;\;\;\; {\rm unless}\;m^\prime=m+1\\
  \langle  \ell^\prime, m^\prime|V_-|\ell, m\rangle &=0 \;\;\;\;\;\; {\rm unless}\;m^\prime=m-1.
 \end{align}
Furthermore the matrix elements vanish unless $\ell' = \ell - 1$ or $\ell' = \ell$ or $\ell' = \ell +1$.
 
Consider the case $\ell' = \ell$. Generalization of the usual arguments
shows that
\begin{eqnarray}
\langle \ell, m+1 | V_+ | \ell, m \rangle & = & A ( \ell - m )^{1/2} ( \ell + m + 1 )^{1/2} 
\hspace{5mm}
m = - \ell, \ldots, \ell - 1 \nonumber \\
\langle \ell, m | V_z | \ell, m \rangle & = & A m 
\hspace{43mm}
m = -\ell, \ldots, \ell \nonumber \\
\langle \ell, m -1 | V_- | \ell, m \rangle & =  & A ( \ell - m +1 )^{1/2} ( \ell + m )^{1/2} \hspace{5mm}
m = - \ell + 1, \ldots, \ell
\nonumber \\
\label{eq:well}
\end{eqnarray}
where the proportionality constant $A$ is a complex number called the
``reduced matrix element''. Note that for ${\mathbf V}$ hermitian,
$A$ would have to be real, but there is no such restriction in the
non-hermitian case.

Similarly in the case $\ell' = \ell + 1$ we find
\begin{eqnarray}
\langle \ell + 1, m + 1 | V_+ | \ell, m \rangle & = & 
B \left[ \frac{ (\ell + m + 2)( \ell + m + 1) }{(2 \ell + 2)(2 \ell + 1)} \right]^{1/2} \nonumber \\
\langle \ell + 1, m | V_z | \ell, m \rangle & = & 
- B \left[ \frac{ (\ell - m + 1) ( \ell + m + 1) }{ (2 \ell + 2) (2 \ell + 1) } \right]^{1/2} \nonumber \\
\langle \ell + 1, m -1 | V_- | \ell, m \rangle & = & 
- B \left[ \frac{ (\ell - m + 1) ( \ell - m + 2 ) }{ (2 \ell + 2) ( 2 \ell + 1 ) } \right]^{1/2}, \nonumber \\
\label{eq:wellplus}
\end{eqnarray}
where $m = - \ell, \ldots, \ell$ and $B$ is another complex reduced
matrix element.

Finally in the case that $\ell' = \ell - 1$ we find
\begin{eqnarray}
\langle \ell - 1, m + 1 | V_+ | \ell, m \rangle & =  &
- C \left[ \frac{ ( \ell - m - 1)( \ell - m ) }{ (2 \ell) (2 \ell - 1) } \right]^{1/2} \nonumber \\
\langle \ell - 1, m | V_z | \ell, m \rangle & = & 
- C \left[ \frac{ (\ell - m)(\ell + m) }{ (2 \ell)( 2 \ell - 1) } \right]^{1/2} \nonumber \\
\langle \ell - 1, m - 1 | V_-  | \ell, m \rangle & = & 
C \left[ \frac{ (\ell + m)( \ell + m - 1 ) }{ (2 \ell)(2 \ell - 1) } \right]^{1/2},
\label{eq:wellminus}
\end{eqnarray}
where $C$ is a complex reduced matrix element and $m = - \ell, \ldots, \ell-2$
in the first line of eq (\ref{eq:wellminus}),
$m = - \ell + 1, \ldots, \ell - 1$ in the
second line of eq (\ref{eq:wellminus}), and $m = - \ell + 2, \ldots, \ell$ in the
last line of eq (\ref{eq:wellminus}).

In the hermitian case the reduced matrix elements satisfy
$B = C^\ast$ but in the non-hermitian case there is no such
restriction on the complex elements $B$ and $C$.


\begin{thebibliography}{99}
\bibitem{bender:1998ke} Bender, Carl M. and Boettcher, Stefan. ``Real spectra in nonHermitian Hamiltonians having PT symmetry'', {\em Phys. Rev. Lett.} {\bf 80},5243-5246 (1998).
\bibitem{bender:1998gh}Bender, Carl M. and Boettcher, Stefan and Meisinger, Peter. ``PT symmetric quantum mechanics'', {\em J.Math.Phys} {\bf 40}, 2201-2229. (1999).
\bibitem{bender:2007nj} Bender, Carl M.,``Making sense of non-Hermitian Hamiltonians",{\em Rept.Prog.Phys.},
{\bf 70}, 947. (2007). 
\bibitem{guo:2009aa} Guo, A. and Salamo, G. J. and Duchesne, D. and Morandotti, R. and Volatier-Ravat, M. and Aimez, V. and Siviloglou, G. A. and Christodoulides, D. N. ``Observation of $\mathcal{P}\mathcal{T}$-Symmetry Breaking in Complex Optical Potentials'', {\em Phys. Rev. Lett.} {\bf 103},093902 (2009). 
\bibitem{ruter:2010ce} Ruter, Christian E. and Makris, Konstantinos G. and El-Ganainy, Ramy and Christodoulides, Demetrios N. and Segev, Mordechai and Kip, Detlef. ``Observation of parity-time symmetry in optics'', {\em Nat. Phys.} {\bf 6} 1515. (2010). 
\bibitem{zhao:2010kf} Zhao, K. F. and Schaden, M. and Wu, Z. ``Enhanced magnetic resonance signal of spin-polarized Rb atoms near surfaces of coated cells'',{\em Phys. Rev. A}, {\bf 81}, 042903 (2010).
\bibitem{benderkalveks} Bender, Carl and Kalveks, R.``$\mathcal{P}\mathcal{T}$" Symmetry from Heisenberg Algebra to E2 Algebra'', {\em Int. J. Theo. Phys},{\bf 50}, 955-962, (2011). 
\bibitem{Coleman:1974jh} Coleman, S. , R. Jackiw and H.D. Politzer, Phys. Rev. {\bf D10}, 2491 (1974). 
\end{thebibliography}
\end{document}